\documentclass[
  ,final            
  ]
  {aipproc}

\layoutstyle{8x11double}

\begin{document}

\title{The Blazhko Effect}



\author{G. Kov\'acs}{
  address={Konkoly Observatory, P.O.Box 67, H-1525, Budapest, Hungary}
}



%
%
\begin{abstract}
Current status of (the lack of) understanding Blazhko 
effect is reviewed. We focus mostly on the various components 
of the failure of the models and touch upon the observational 
issues only at a degree needed for the theoretical background. 
Attention is to be paid to models based on radial mode 
resonances, since they seem to be not fully explored yet, 
especially if we consider possible non-standard effects (e.g., 
heavy element enhancement). To aid further modeling efforts, 
we stress the need for accurate time-series spectral line analysis 
to reveal any possible non-radial component(s) and thereby 
let to include (or exclude) non-radial modes in explaining the 
Blazhko phenomenon.   
\end{abstract}

\maketitle

%
%
\section{Introduction}
Efforts in understanding the Blazhko effect (periodic amplitude-  
and phase-modulation of the pulsation of RR~Lyrae stars -- 
\cite{bla1907}, \cite{sha1916}) date back nearly to the discovery 
of their variability itself \cite{bai1899}. Unfortunately, the 
situation has not changed much since then. We are stunned by the 
overall simplicity of the phenomenon and the lack of theory that 
is able to account for the basic observed properties. Although 
several observational facts indicate that the theory of horizontal 
branch evolution and RR~Lyrae pulsation is basically correct, and 
it is unlikely that the future understanding of the Blazhko phenomenon 
will involve dramatic changes in these models, it is still highly 
unacceptable that we do not fully understand this basic class of 
variables. We agree with \cite{dzi2004}:

\medskip 
{\em ``Explaining periodic, or nearly periodic, long term 
modulation of the light curve seen in many RR~Lyraes 
-- known as Blazhko effect -- constitutes perhaps the 
greatest challenge to stellar pulsation theory.''} 
\medskip 

In this review we briefly summarize the various efforts that have 
taken place since similar reviews of the topic (\cite{kov2002}, see 
also \cite{smi2006} and \cite{kol2008}). We attempt to give a simple 
discussion of the reasons leading to the failure of each of the 
current models. As a matter of curiosity we also present a toy model 
that, for the first time, shows a behavior similar to what we observe 
in Blazhko stars. We conclude with suggested directions of research 
and expectations from satellite data.  

%
%
\section{Observational status -- a quick glance}
\subsection{RR~Lyrae stars}
There have been several thorough works done during the past seven 
years that have substantially contributed to establishing a more 
complete picture on the observed properties of Blazhko-type pulsation. 
In addition to the analyses of the large photometric surveys of MACHO, 
OGLE and ASAS \cite{alc2003}, \cite{sos2008}, \cite{szc2007}, there 
are observational projects focusing on the particular problem of 
Blazhko modulation. Johanna Jurcsik and coworkers at the Konkoly 
Observatory have been monitoring these objects photometrically by 
a dedicated 60-cm telescope since 2004. ``The Blazhko Project'' 
led by Katrien Kolenberg at the University of Vienna studies specific 
targets both by photometric and by spectroscopic means. Further 
photometric studies are made at the Michigan State University by 
Horace Smith and his students. Very importantly, there are 
spectroscopic investigations carried out by Merieme Chadid, Katrien 
Kolenberg and coworkers aiming at to measure magnetic fields in  
RR~Lyrae stars. We refer to the following home pages on the works 
done by these groups.\footnote{ 
http://www.konkoly.hu/24/

\hspace{-1.4mm}
http://www.univie.ac.at/tops/blazhko/

\hspace{-1.4mm}
http://www.pa.msu.edu/astro/smith/
}

The main photometric properties of the Blazhko stars are very simple 
(as compared to some familiar classes of multimode variables, e.g., 
to those of the $\delta$~Scuti stars). The frequency spectra of the 
light curves are (usually strongly) dominated by the pattern of 

\medskip
\fbox{$\nu_{\rm BL}$, \{$k\nu_0$\}, \{$k\nu_0\pm\nu_{\rm BL}$\}} 
\medskip

\noindent
where $\nu_0$ is the main pulsational frequency (that of the fundamental 
[FU] or the first overtone [FO] mode), $\nu_{\rm BL}$ is the modulation 
(the Blazhko) frequency and $k=1,2,...$, usually goes to fairly high 
harmonics in the case of the FU (RRab) variables. The amplitudes of the 
modulation components corresponding to \{$\pm\nu_{\rm BL}$\} are usually 
asymmetric: 

\medskip
$A(k\nu_0-\nu_{\rm BL})\not = A(k\nu_0+\nu_{\rm BL})$
\medskip

\noindent
The highly asymmetric cases often go to the extreme, with one side 
under the detection limit -- e.g., \cite{ole1999}. These cases lended 
further support to the non-radial mode excitation concept. At the 
same time, the explanation of this asymmetric amplitude pattern 
(even if it is only a mild one) is the main stumbling block of the 
current models invoking non-radial pulsations.  

In a non-negligible number of cases the frequency pattern is 
more complicated, starting from the appearance of higher-order 
modulation side lobes (see \cite{juh2009}, \cite{kog2009}, 
\cite{hur2008}) and ending with more complex structures around 
the pulsation components (e.g., \cite{wil2008}, \cite{nag2006}, 
\cite{jur2008}) or even more intriguingly, with the appearance 
of very low-amplitude structures with the same modulation 
frequency but centered at `no-clue' frequencies (see Chadid et al., 
these Proceedings). All these add to the flavor to the Blazhko 
phenomenon and could be instrumental at some stage in searching 
for the physical model, but the more general pattern described 
above constitutes the observational framework the models should 
deal with. 

%
%
\medskip
\begin{table}[h]
\begin{tabular}{lcccc}
\hline
\tablehead{1}{c}{}{Type/Field\tablenote{\underline{Sources:}
\cite{sos2002}(SMC), \cite{alc2003}(LMC), \cite{miz2004}(SMC), 
\cite{nag2006}(LMC), \cite{col2006}(Bulge), \cite{szc2007}(Gal. Field), 
\cite{mos2008}(compilation)}}
  & \tablehead{1}{c}{c}{LMC}
  & \tablehead{1}{c}{c}{SMC}
  & \tablehead{1}{c}{c}{Gal. Bulge}
  & \tablehead{1}{c}{c}{Gal. Field\tablenote{\underline{Note:}
 See comments in text for the expected increase of the incidence 
 rates for Galactic RR~Lyrae stars and for the estimation of the 
 rates in the case of multiple sources for the statistics.
 }}   \\
\hline
RRab & $12$\% & $10$\% & $25$\%  & $>5$\% \\
RRc  & $10$\% & $10$\% & $12$\%  & $>7$\% \\
\hline
\end{tabular}
\caption{Statistics of Blazhko stars.}
\label{table1}
\end{table}  

The relative number of Blazhko stars varies in different stellar 
populations. Table~1 attempts to give a current census. When multiple 
sources are available, we present the values considered to be the 
`most reliable'. We caution that considerable changes can be expected 
in certain populations due to the advance of measurement and data 
analysis techniques. This is especially true for the Galactic field 
population where both the ground-based survey of Jurcsik et al. 
(these Proceedings) and analysis of a handful of CoRoT RR~Lyrae stars 
by Szab\'o et al. (these Proceedings) suggest that the incidence rate 
could be (even) higher than $50$\%. Obviously this is an important 
issue from the point of view of model builders and it will be interesting 
to see the distribution of the modulation amplitudes; whether they 
follow a continuous long tail to the micro-modulation regime or they 
show a bimodality with peaks at the `classical', high-amplitude Blazhko 
modulation and in the micro-modulation range.

\subsection{Cepheids}
Before 2004, except for V473 Lyrae (HR~7308, see \cite{bur1982} 
for the analysis of the radial velocity and \cite{koe2001} of 
the Hipparcos data) no Cepheid variable was known with modulated 
light variation, similar to those of the Blazhko stars. The 
analysis of the OGLE database of the LMC by Moskalik et al. 
\cite{mos2004} has led to the discovery of a substantial 
number of Blazhko-type Cepheids. This is a very significant 
result, since then we need to seek for models of similarly 
modulated pulsations in two classes of stars that are dissimilar 
to a great extent. In Table~2 we summarize the current statistics 
on these stars, based on two available sources \cite{sos2008} and 
\cite{mos2009}. The two studies yield basically consonant statistics, 
in spite of the different datasets the works are based on 
(\cite{sos2008} use mostly OGLE III, that is more extended and more 
accurate than those of OGLE II of \cite{mos2009} -- however, 
\cite{sos2008} do not give any details of the analysis but only 
the final statistics). The systematically lower rates in 
\cite{mos2009} can be attributed to the less extensive data used 
in their analysis.  

%
%
\medskip
\begin{table}[h]
\begin{tabular}{lrrrrr}
\hline
\tablehead{1}{c}{}{Type/Ref}
  & \tablehead{1}{c}{c}{FU}
  & \tablehead{1}{c}{c}{FO}
  & \tablehead{1}{c}{c}{SO}
  & \tablehead{1}{c}{c}{FU/FO}
  & \tablehead{1}{c}{c}{FO/SO}   \\
\hline
\cite{sos2008}  & $4$\% & $28$\% & $28$\%  & $18$\% &  $36$\% \\
\cite{mos2009}  & $0$\% &  $8$\% & $12$\%  & $13$\% & $>19$\% \\
\hline
\end{tabular}
\caption{Statistics of modulated Cepheid variables in the LMC.}
\label{table2}
\end{table}  

We see that modulations are observable in all types of Cepheids 
in quite high rates. RR~Lyrae stars are apparently different 
in this respect. In the case of double-mode (FU/FO) RR~Lyrae 
stars the Blazhko behavior has been reported only in one 
object \cite{nag2006}. Interestingly, the phenomenon 
seems to be quite common among double-mode Cepheids, especially 
among the FO/SO ones. It is important to note that the average 
of the modulation level $A_{\rm mod}/A_{\rm puls}$ of Cepheids 
is only about $5$\%, i.e., substantially lower than those of 
the RR~Lyrae stars -- e.g., \cite{szc2007}. Except for the FO/SO 
stars, which have modulation periods greater than $>700$~days, 
all other types of Cepheids span a similar range of modulation 
periods as those of the RR~Lyrae stars (compare \cite{alc2003} 
and \cite{mos2009}). A very interesting feature of the modulated 
FO/SO variables is the opposite phase by which the amplitudes of 
the FO, SO modes change (in the modulation interpretation of the 
observed frequency pattern). This is certainly at odds with models 
based on geometric effects in explaining amplitude modulation (i.e., 
the oblique rotator and the 1:1 resonance models). Some dynamical 
mode energy exchange model might be more appropriate (as mentioned 
in \cite{mos2009}), but there is no idea on the specific physical 
and mathematical details of this interaction. In any case, these 
new discoveries may perhaps open new ways in searching for additional 
ideas/models of modulated stellar pulsations.

%
%
\section{Current models -- why do they fail?}
In this section we briefly describe various ideas to explain 
modulated stellar pulsations and then discuss in some detail 
the three, currently most often cited models/ideas. We stress 
that there are very important differences in the level of 
elaboration among these models.  

In terms of elaboration the non-radial resonant rotator/pulsator 
(NRRP) model stands out, since it is based on the detailed 
analysis of the {\em nonlinear} interaction among radial and 
non-radial modes (see \cite{van1993} and \cite{gou1994}). 
We recall that nonlinearity is a basic attribute of the Blazhko 
phenomenon, since it is observable throughout many decades, 
i.e., well above the dynamical (or even thermal) time scales. 
As far as the methodology is concerned, the drawback (and, 
because of its generality, as well as the power) of the NRRP 
model is that it is based on the amplitude equation (AE) formalism 
(see \cite{dzi1980}, \cite{dzi1982} and 
\cite{buc1984})\footnote{Due to the extreme level of complexity 
of 3D time-dependent long-term hydrodynamical simulations, 
the only method currently available to us to study nonradial 
nonlinear pulsations is to use AEs.}, and, as a result, no strong 
specific predictions can be made for the observed behavior of 
particular stars. Nevertheless, the predictive power of the 
NRRP model is not worse than any other available models/ideas.        

The magnetic oblique rotator/pulsator (MORP) model lacks the 
complexity due to nonlinearity and does not examine the stability 
of the presumed non-aligned magnetic, rotating and pulsating 
configuration. Instead, it postulates the stability (perhaps based 
on the observations of spotted stars and rapidly oscillating Ap 
stars) and gives observed properties in a pure {\em kinematic/linear 
pulsation} framework. 

The 2:1 resonance model of \cite{mos1986} is also based on the AEs 
and assumes radial mode interaction only. This is the first (and 
so far the only) model that represents the modulation by dynamical 
mode interaction. It is reasonably elaborated (further works were 
probably discouraged by the lack of hydrodynamically confirmed 
numerical computations). In this review we construct hydrodynamical 
models that may suggest some level of viability of this idea. 

There are also ideas that consider some sort of energy transfer 
hypothesis among some pulsation modes to aid amplitude modulation. 
The hypothesis of \cite{dzi2004} assumes that there is an energy 
sink originating from many low-amplitude nonradial modes. The 
interaction leads to lowering the amplitude of the radial modes, 
that eventually results in a modulation of the pulsation (through 
some, so far unspecified interaction between these two classes of 
modes). The authors put forward the slight effect of the overall 
lower amplitude of the Blazhko stars in the Galactic Bulge as compared 
to that of the single-period stars (see also the exhibition of the 
same effect for the Galactic field and M3 variables in \cite{sze1988} 
and for the LMC in \cite{alc2003}). However, a simple resonant mode 
interaction can also lead to amplitude decrease (see \cite{buc1986}) 
and there is no guarantee (or any obvious physical reason) that such an 
interaction will lead to modulated pulsation. The argument by 
\cite{mos2009} for a possible dynamical interaction in explaining 
the amplitude modulations of the FO/SO double-mode Cepheids is more 
convincing, since the data can indeed be interpreted as the result 
of such an interaction. However, the authors stay vague in specifying 
the type of interaction that may take place in these objects. 

The idea of multimode non-resonant interaction also came up in 
\cite{kov1994} as a possible source of amplitude modulation (for 
the analysis of the three-mode non-resonant interaction see also 
\cite{ish1990} and \cite{ver1990}). On general physical ground, 
it is easy to show that more than two modes are necessary to expect 
non-stationary (i.e., non-constant amplitude) asymptotic 
states.\footnote{Stationary 
(constant amplitude) states correspond to steady multimode pulsations. 
These states are not interesting if only non-resonant radial modes are 
considered, because they are sparsely spaced and do not give rise to 
long-term modulation.} 
Although an extensive parameter search may lead to finding modulated 
solutions, it is nearly impossible to construct one in which one mode 
strongly dominates the others (as observed in real stars).   

In a current idea presented in \cite{sto2006} we find a loose 
description of the highly complex scenario of magnetic/convective 
interaction with radial pulsation as a possible cause of modulation. 
Although we cannot exclude that such a complicated system may also 
lead to the modulation of radial pulsation, there are some 
misinterpretations present in that paper and in concomitant incorrect 
applications. These will be discussed in a separate subsection below.  
 
%
%
\subsection{General comments on the relevance of AEs}
Before dwelling upon the more detailed discussion of the various models 
of modulated pulsation, we recall some of the relevant properties of 
the AEs. Most part of the subject discussed below can be summarized 
simply as a warning against the straightforward application of the 
linear and nonlinear asymptotic pulsation results to the various 
pulsation states of the Blazhko modulation. Furthermore, the general 
framework of AEs enables us to test the ability of any new effect to 
cause modulated states. 

First of all, the concept of AEs is very general and can be applied 
whenever we would like to study the basic {\em nonlinear} properties 
of a dynamical system where the interaction goes on largely different 
time scales. For example, we can mention the long-term perturbations 
in celestial mechanics \cite{kev1970} or processes on time scales 
considerably longer than the dynamical time scales in pulsating stars 
(such as mode switching or amplitude modulation, see \cite{dzi1980}, 
\cite{dzi1982}, \cite{buc1984}). Secondly, in many cases AEs are the 
{\em only} means to study complicated systems, where numerical 
simulations are either too expensive or impossible (e.g., nonlinear 
nonradial stellar pulsations). 

An important underlying assumption in deriving AEs for a given 
dynamical system, is that the nonlinearities are {\em mild}, 
meaning that they can be represented as a Taylor expansion in 
the neighborhood of the linear regime (that is usually rather 
easy to describe fairly accurately due to the separability of 
the time and spatial variations, and because of the linear 
superposition of eigenmodes in the general solution). If the 
above conditions are satisfied, then we start off by the following 
form of some physical quantity, e.g., of the displacement vector:

%
%
\begin{eqnarray}
\delta\vec{r}(\vec{s},t) & = & a_0(t)\vec{\xi_0}(\vec{s}) + a_1(t)\vec{\xi_1}(\vec{s}) + ..., \hskip 2mm ,
\end{eqnarray}

\noindent
where $\vec{s}$ is the spatial coordinate, $\vec{\xi_k}$ is the linear 
non-adiabatic eigenvector of the normal-mode `k' and $a_k(t)$ is the 
corresponding time-varying amplitude (complex, as all quantities in 
Eq.~1). The time-dependent functions are factorized to slow- and and 
short-varying parts and the AEs are derived for the slow-varying part. 
When Eq.~(1) is substituted in the original partial differential 
equations that describe the fluid dynamics, we omit the short-oscillating 
terms and derive the AEs by demanding boundedness for the solution. 

The most important part of this process from the present context is that 
we start from the {\em normal mode spectrum}. This and the resonances 
will determine the basic structure of the AEs representative for a given 
system. For example, if we consider the radial pulsation of a fully 
radiative stellar envelope in the gray atmosphere approximation, we need 
to solve a system that is 3rd-order in time (two orders come from the 
time-dependence of the shell motion and another one comes from the 
entropy variation [see, e.g., \cite{buc1990}]). Therefore, this system 
leads to a $3\times N$-order eigenvalue problem for the normal mode 
frequencies ($N$ stands for the number of mass shells). Since the matrix 
of this eigenvalue problem constitutes of real elements, the eigenvalues 
are either pure real or come in complex conjugate pairs and a set of 
real ones. Due to the underlying physical setting, all radiative stellar 
models for classical pulsators display this latter type of spectrum. 
The non-oscillating modes (due to their strong damping and different 
spatial variation) are usually considered to play no role in the 
behavior of the oscillating modes. The oscillating non-resonant normal 
mode spectrum leads to the following set of AEs in the lowest order 
approximation: 

%
%
\begin{eqnarray}
{dA_k\over{dt}}       & = & A_k(\kappa_k+Q_{k0}A_0^2+Q_{k1}A_1^2+...) \\
{d\varphi_k\over{dt}} & = & R_{k0}A_0^2+R_{k1}A_1^2+... \hskip 2mm ,
\end{eqnarray}

\noindent
where, $A_k(t)$ and $\varphi_k(t)$ stand for the slow-varying part of 
$a_k(t)$ in Eq.~(1), i.e., the time-varying part of the mode behavior 
is described by $A_k(t)\sin(\omega_kt+\varphi(t))$, with $\omega_k$ 
corresponding to the linear non-adiabatic eigenfrequency and $\kappa_k$ 
to the linear growth rate. The coefficients \{$Q_{ki}$\} and \{$R_{ki}$\} 
are the non-resonant {\em coupling coefficients}. These are complicated 
expressions of the eigenfunctions and stellar structure, and rarely 
attempted to compute {\em ab initio} (see, however, \cite{van1993}). 
On the other hand, simple physical considerations and numerical models 
show that each member of the set \{$Q_{ki}$\} is {\em negative}. 
Similar statement may hold also for \{$R_{ki}$\} -- i.e., \cite{buc1987}. 

It follows that for single-mode non-resonant pulsation, corresponding 
to the {\em asymptotic} state, after dying out the transients due to, 
e.g., the interaction with the other modes, the amplitude and the 
{\em nonlinear} period can be written in the following form:      
   
%
%
\begin{eqnarray}
A_k^2   & = & -{\kappa_k\over{Q_{kk}}} + {\rm H.O.T.} \\
P_k(NL) & = & {2\pi\over{\omega_k + R_{kk}A_k^2}} + {\rm H.O.T.}\hskip 2mm ,
\end{eqnarray}

\noindent
where H.O.T. stands for `higher order terms' (in amplitude). Since 
the coupling coefficients are slowly varying functions of the stellar 
parameters, the global properties of the single-mode non-resonant 
pulsation can be approximated from the behavior of the linear growth 
rates.  

With these preliminaries, in the context of this review, we think that 
attempts trying to relate the observed temporal properties of Blazhko 
stars in terms of asymptotic and linear pulsation results are {\em improper} 
applications, because of the following two important reasons.
\begin{itemize}
\item
Variations that take place in the large-amplitude asymptotic regime 
of stellar pulsation (i.e., in the observed variables) are related to 
the above (but not the same) types of equations and {\em not} to the 
ones described by the infinitesimal oscillations around the static 
state (i.e., the linear pulsational relations, e.g., van Albada 
and Baker \cite{van1973}).
\item
Equations (4) and (5) are {\em irrelevant} from the point of view of 
the Blazhko effect, since they are asymptotic results whereas the time 
scale of the Blazhko effect is, in general, much shorter than the 
relaxation time of the perturbations near the limit cycles.\footnote{The 
actual time scale depends on the type of perturbation and on the model 
(see, e.g., \cite{sza2004}), but for stable, dominantly single-mode 
pulsation in overtone `k' this time scale is of the order of 
$\kappa_k^{-1}$.} Furthermore, even if the time scales are comparable, 
application of the asymptotic results (e.g., the nonlinear period 
change, see \cite{sto2006}, \cite{jur2009}) at various phases of the 
modulation, is in obvious contradiction with the original meaning of 
the above asymptotic values, since these kinds of applications assume 
instantaneous adjustment to the asymptotic state described by the 
non-modulated model. 
\end{itemize}

Based on the tight relation of the eigenspectrum of the linear 
problem and the type of AEs we can deal with in searching for 
systems exhibiting amplitude modulation, new/better physics 
can yield such a system, if they generate new modes that can 
interact with the radial pulsation modes. In principle, 
convection might be a good candidate for generating such 
``new modes'', but, so far,  there are no reports on the 
instrumental nature of such modes, although current 1D 
convection/pulsation models do include several important 
properties of convection (\cite{ste1982}, \cite{buc1999}, 
\cite{smo2008}, see also Buchler, these Proceedings). We do 
not see what kind of new treatment of convection can lead to 
a convective eigenmode spectrum that could give rise to AEs 
capable of leading to amplitude modulation (see some additional 
notes on the hypothetical role of resonant convective modes in 
\cite{kov1994}).

%
%
\subsection{Magnetic oblique rotator/pulsator}
The first mentioning of an oblique-rotation magnetic/pulsating 
configuration as a possible explanation of the the Blazhko 
phenomenon comes from \cite{bal1959} and \cite{det1973}. Although 
subsequent magnetic measurements by \cite{bab1958} and \cite{pre1967} 
were controversial, the idea was further kept as a possible 
explanation due to some long-term modulation superposed on the 
41-day Blazhko cycle of RR~Lyrae as being the manifestation of 
some magnetic star cycle, analogous to the solar cycle \cite{sze1976}. 
Recent measurements of magnetic field by \cite{cha2004} strongly 
suggest that RR~Lyrae itself should have a field lower than 
$\sim 100$~G. A similar upper limit was obtained by \cite{kol2009} 
by the spectropolarimetric survey of $17$ galactic field RR~Lyrae 
stars that included at least $10$ Blazhko stars. 

The detailed modeling of the oblique rotator idea was largely 
motivated by the success of the same type of model by Kurtz 
\cite{kur1982} explaining the amplitude variation of the 
nonradial modes of the rapidly oscillating Ap stars. The first 
MORP model for the Blazhko effect by \cite{cou1983} was put in 
a more strict theoretical framework by \cite{tak1994} and \cite{tak1995} 
as summarized in \cite{shi2000}. The essential result in these 
{\em purely linear} works is that under large-scale magnetic 
fields {\em all modes} become {\em nonradial}. More specifically, 
it was shown that in the case of a pure dipole magnetic field the 
eigenmode corresponding to a radial mode in the non-magnetic case 
will be {\em always} deformed spatially by an $l=2$ spherical 
harmonic component. It follows that in the simplest setting we 
expect a quintuplet frequency structure around the radial 
pulsation component. The nice feature of this model is that 
the modulation components are scaled by the magnetic field 
strength and geometric factors (i.e., inclination angles) and 
these can be determined from the observed amplitude spectrum. 
The two important obstacles that jeopardize this model are the 
following. 
\begin{itemize}
\item
It follows from the above that the MORP model is purely geometric 
and the modulation is caused by rotation. Therefore we have 
{\em strictly symmetric} quintuplets both in amplitudes and in 
frequency separations. There is no easy way to produce asymmetric 
modulation components in the frequency spectra.\footnote{In \cite{shi2000} 
there is a note on a possible break of this symmetry by an off axis 
dipole magnetic field, however, no detailed elaboration is given.} 
\item 
The required strength of the dipole field is of the order of $1$~kG. 
This is an order of magnitude larger that the one suggested by current 
measurements (see above).      
\end{itemize}

%
%
\subsection{Nonradial resonant rotator/pulsator}
For an early, brief discussion and references on the possibility that 
nonradial modes might be instrumental in the Blazhko effect we 
refer to \cite{ste1976}. Starting in the early nineties, the idea 
has been revisited on a more fundamental basis by using updated 
linear numerical models and AEs (van Hoolst and coworkers in 
\cite{van1995} and \cite{van1998}, see also \cite{kov1994} for 
a brief analysis of the relevant EAs and \cite{gou1994} and 
\cite{buc1997} for the discussion of AEs involving various 
nonradial modes). Further thorough elaboration of these ideas 
and specific extension to the Blazhko stars has been made by 
Nowakowski and Dziembowski in \cite{now2001} (see also 
\cite{dzi1999}).     

In the NRRP model we assume that the radial mode of interest 
(the FU or FO modes) are in 1:1 resonance with one of the 
not too strongly damped nonradial modes. Based on the models 
of \cite{van1998} and \cite{now2001}, we can strongly 
suspect that there are many nonradial modes of $l=1$ that are 
in close resonance with the mode of interest and they are among 
the ones that are most viable for excitation (even though the 
code, on which this conjecture is based, does not include the 
effect of the interaction between pulsation and convection). 
The large number of available nonradial modes comes from the 
fact that in evolved stars the nonradial modes have high-order 
g-mode character in the deep interior. On the other hand, in 
the outer part of the stellar envelope they are similar to 
the p-modes and therefore, they have some chance to be spatially 
coupled with the radial modes (an important property when mode 
interaction is considered).  

From these encouraging linear properties it is suspected that 
the relevant AEs describing the resonant interaction between 
the radial and the nonradial mode(s) may lead to some stable 
state and thereby giving a chance for amplitude modulation either 
through some geometric or true dynamical interaction between 
the two classes of modes. In concentrating on the most viable 
interaction between the radial and the $(l=1,m=0)$ or  
$(l=1,m=\pm1)$ nonradial modes, it was shown by \cite{now2001} 
that, in the asymptotic regime (i.e., for large time scales), the 
two types of interaction can be analyzed with the same set 
of equations valid for the 1:1 resonance with $(l=1,m=0)$. 
Furthermore and very importantly, the most probable solution 
of the relevant AEs is the {\em stationary} one, with 
{\em constant amplitudes and relative phase}. This means that 
the interaction with the $(l=1,m=0)$ mode {\em does not} 
lead to amplitude modulation, because of the constant 
relative phase. This results in the {\em phase lock} phenomenon 
(see \cite{dzi1984}), implying frequency synchronization, which, 
in the present case, means monoperiodic pulsation. Therefore, 
the NRRP model relevant for the Blazhko effect should consist 
of a 1:1 resonant stationary interaction with the $(l=1,m=\pm1)$ 
modes and due to phase lock and stellar rotation, we observe a 
modulated pulsation. Although it is described briefly in 
\cite{now2001}, and it is a simple derivation, we think that 
it is useful to see how the final frequency separation is related 
to the stellar rotation. 

Let the stationary {\em nonlinear} frequencies of the $(l=1,m=-1)$, 
$(l=0)$ and $(l=1,m=+1)$ modes, respectively, in the co-rotating 
frame, be $\omega_{-}$, $\omega$ and $\omega_{+}$. We adorn the 
corresponding {\em linear} values by zero superscript. Then, 
denoting the corresponding nonlinear changes by $\delta\omega_{-}$, 
$\delta\omega$ and $\delta\omega_{+}$, we have:  
%
%
\begin{eqnarray}
\omega=\omega^0+\delta\omega,\hskip 2mm 
\omega_{\pm}=\omega_{\pm}^0+\delta\omega_{\pm} \hskip 2mm . 
\end{eqnarray}
If the frequency distances between the central peak (corresponding 
to the radial mode) and the side peaks (corresponding to the nonradial 
modes) are denoted by $\Delta_{\pm}$, we get: 
%
%
\begin{eqnarray}
\Delta_{-}&\equiv&\omega - \omega_{-}=
\omega^0 + \delta\omega - \omega_{-}^0 - \delta\omega_{-} \\
\Delta_{+}&\equiv&\omega_{+} - \omega=
\omega_{+}^0 + \delta\omega_{+} - \omega^0 - \delta\omega \hskip 2mm . 
\end{eqnarray}
With the {\em linear frequency mismatch} of 
$\Delta\omega=2\omega_0-\omega_{-}^0-\omega_{+}^0$, it follows that 
the difference between the two spacings can be expressed as:   
%
%
\begin{eqnarray}
\Delta_{+}-\Delta_{-}=
-2\Delta\omega+2\delta\omega-\delta\omega_{-}-\delta\omega_{+} \hskip 2mm . 
\end{eqnarray}
Then, using the stationarity condition and comparing the right-hand-side 
of the above expression with Eqs. (23) and (31) of \cite{now2001}, we 
see that it is equal to zero, i.e., we have an equidistant triplet in 
the final state. Since the $m=\pm1$ modes have the same radial structure, 
the corresponding nonlinear frequency changes are also the same, 
i.e., $\delta\omega_{-}=\delta\omega_{+}$. Therefore, the frequency 
separations can be expressed as follows: 
%
%
\begin{eqnarray}
\Delta_{-}&=&(\omega^0 + \Delta\omega) - \omega_{-}^0 \\
\Delta_{+}&=&\omega_{+}^0 - (\omega^0 + \Delta\omega) \hskip 2mm . 
\end{eqnarray}
Thus, the frequency separation $\Delta$ can be computed from the 
classical Ledoux formula, because: 
%
%
\begin{eqnarray}
\Delta={\Delta_{-}+\Delta_{+}\over {2}}=
{\omega_{+}^0-\omega_{-}^0\over {2}} \hskip 2mm . 
\end{eqnarray}
With $\Omega$ rotational angular velocity we have: 
$\omega_{\pm}^0=\omega_l^0\pm C\Omega + D\Omega^2 + ...$. Since 
the nonradial modes are high-order g-modes, the coefficient $C$ 
is equal to $0.5$ in the asymptotic limit for $l=1$ modes. The 
exciting outcome of all these is that if this model were viable, 
then the Blazhko period would be directly related to the rotational 
period of the star, so nonlinear period change would not influence 
the simple frequency splitting that we naively expect. In addition, 
the corresponding rotational rates are not in contradiction with the 
spectroscopically derived values \cite{pet1996}, although for very 
short Blazhko periods, we might have problems \cite{jur2005}. There 
are also variables with multiple modulation periods \cite{wil2008}. 
Obviously, the interpretation of these objects in terms of rotation 
modulation is even more difficult. 

As far as the amplitudes of the modulation components are concerned, 
they will be symmetric both for physical and for geometrical reasons 
(same type of spatial dependence and rotation). As a result, the 
main reason of the fall of the NRRP model can be summarized as 
follows. 

\begin{itemize}     
\item
In the basic model described above the modulation components have 
strictly {\em equal amplitudes}. This property is in clear contradiction 
with the observations that show the overwhelmingly opposite behavior 
of Blazhko stars. Perhaps the consideration of additional nonradial 
modes could lead to asymmetric amplitudes, but this would require 
the introduction of additional unknown parameters and the result 
would not be verifiable.   
\end{itemize}     

However, there seems to be also a deeper problem with our current 
understanding of nonradial mode coupling in evolved stars. In 
\cite{now2003} the question has been posed if the condition of mild 
nonlinearity (one of the basic assumptions behind the applicability 
of the AE formalism) is satisfied for both the radial and the 
horizontal components of the nonradial modes. The most essential 
part of the reasoning is that the horizontal component of the 
displacement vector is proportional to the Brunt-V\"ais\"al\"a 
frequency (see their Eqs. (1), (2)). This latter quantity is a very 
sensitive function of the internal structure and can reach very 
high values near the core of evolved objects. Figure~1 shows the 
variation of the relative Brunt-V\"ais\"al\"a frequency as a 
function of the distance from the star's center. We see the large 
difference between (near) Main Sequence and Horizontal Branch stars. 
In the latter the Brunt-V\"ais\"al\"a frequency can be higher by nearly 
two orders of magnitude in the deep interior than in MS stars. The 
large value implies also large horizontal displacement (several 
hundred times larger than that of the radial component). Based on 
this result, the authors also make an estimate on the expected size 
of the critical surface amplitudes of the nonradial modes, where 
nonlinearity becomes strong (i.e., advective terms start to dominate 
the fluid motion). For $l=1$ modes in standard RR~Lyrae models they 
obtained an amplitude of $0.02$~mag. This is considerably smaller 
than the generally observed modulation side lobes in the frequency 
spectra of the Blazhko stars. These results suggest that the standard 
AE formalism that assumes mild nonlinearity might not be applicable 
for studying nonradial mode interaction in evolved stars.     

%
%
\begin{figure}[h]
  \includegraphics[height=.33\textheight]{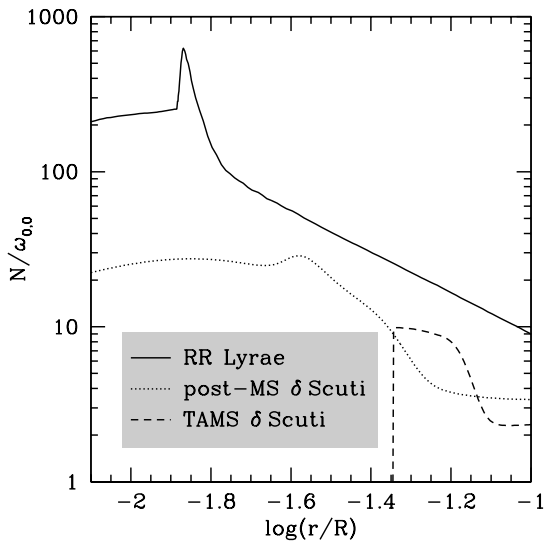}
  \caption{Variation of the Brunt-V\"ais\"al\"a frequency 
           for different stellar models as a function of the 
	   stellar radius (adapted from arXiv:astro-ph/0301504v1, 
	   see also \cite{now2003}).}
\end{figure}

%
%
\subsection{Dynamo+convection+pulsation + criticism}
The purpose of this subsection is to draw attention to the 
inconsistencies occurring in a recent idea of Blazhko effect 
presented in {\cite{sto2006}. Some of the problems concerning 
this idea have already been mentioned in earlier sections. 
Here we give a brief, but more complete review. 

\begin{itemize}
\item
First of all, the idea lacks the level of elaboration observed 
in all other often cited models (MORP, NRRP and 2:1 resonance 
models -- see next section). There is no prediction on the 
expected modulation period or the level of modulation. 
No mentioning is made on the generic case of unequal modulation 
side lobes (one serious problem that makes -- among others -- 
the MORP and NRRP models fail). 
\item
Secondly, there is a misinterpretation of the nonlinear hydrodynamical 
results concerning the relevance of the asymptotic periods in 
explaining the temporal periods appearing in one type of mathematical 
forms of the Blazhko modulation. This topic is discussed already 
earlier, so we just mention that: (a) All nonlinear radially pulsating 
stellar models show small {\em period increase} in their final single-mode 
states (except for \cite{bon1994}, but see \cite{kov1988}, \cite{dek2008} 
and also Smolec, private communication); (b) These period changes have 
nothing to do with the temporal period (as defined by the time derivative 
of the total phase) of the Blazhko modulation. Nonlinear period changes 
are {\em asymptotic} and reaching these states takes a long time; 
(c) The amplitude-dependence of the nonlinear periods as predicted by 
the AEs has also an asymptotic meaning, and cannot be applied to explain 
the possible correlation between the temporal phase (i.e., period) and 
amplitude values of the Blazhko modulation. 
\item
Thirdly, the heart of the idea is an assumed periodic build up and 
destruction of a magnetic field via the interaction with the turbulent 
convection. No mentioning is made on the expected strength of this 
field and why this would lead to an amplitude modulation. The way it 
is presented, sounds like if the assumption was that pulsation is 
working in an environment determined by the interaction between 
convection and turbulence and there is no feedback between them. 
If this is so, then the idea lacks the important effect of 
convection--pulsation interaction, and, due to the assumed magnetic 
field-turbulence interaction, the pulsation--magnetic field feedback 
is also missing. In very general terms, it is unclear why this highly 
complicated system would lead to AEs that are capable of giving 
rise to modulated pulsation. (We recall that the eigenmode spectrum 
of a dynamical system will determine the possible forms of AEs, so 
not knowing the linear properties of the {\em full} linear system, 
we cannot predict what happens in the nonlinear regime). 
\end{itemize}
     
In conclusion we think that the complex idea depicted in \cite{sto2006} 
could be verified only if some sort of deeper physical modeling were 
possible, first in the regime of linear magnetohydrodynamical 
convective/pulsation models. This seems to be quite a complicated 
but perhaps not an impossible task.  

%
%
\section{A resonant toy model -- for entertainment}
This section is by {\em no means} to suggest that the 
model presented could be seriously considered as one of 
the possible explanations of the Blazhko effect. Our goal 
is just to introduce a non-conventional hydrodynamical model 
that shows an amplitude- and phase-modulation, resembling  
to what we see in the observations. Since this is the 
first time such a toy model is presented, we think it 
may be interesting and will perhaps encourage others to 
pursue some other types of non-standard models.  

In 1986 Moskalik \cite{mos1986} suggested that a 2:1 
resonance between the fundamental and third overtone 
modes in RR~Lyrae stars might cause real dynamical 
amplitude/phase modulation. The idea follows a model 
of double-mode pulsation \cite{dzi1984} which was 
thought to be encouraging at that time, and which 
utilizes the same type of resonance. The equations 
governing the mode interaction under this resonance 
are as follows:  
%
%
\begin{eqnarray}
dA_1\over {dt}  & =  & A_1(\kappa_1 + q_{11}A_1^2 + q_{12}A_2^2 + q_{13}A_3^2) \nonumber \\ 
                & +  & C_1A_1A_3\ \cos\Phi \\       
dA_2\over {dt}  & =  & A_2(\kappa_2 + q_{21}A_1^2 + q_{22}A_2^2 + q_{23}A_3^2) \\ 
dA_3\over {dt}  & =  & A_3(\kappa_3 + q_{31}A_1^2 + q_{32}A_2^2 + q_{33}A_3^2) \nonumber  \\
                & -  & C_3A_1^2\ \cos\Phi \\         
d\Phi\over{dt}  & =  & r_1A_1^2 + r_3A_3^2  \nonumber \\               
                & +  & \Delta\omega + \biggl({C_3A_1^2\over{A_3}}-2C_1A_3\biggr)\sin\Phi      
\hskip 2mm .
\end{eqnarray}
Here we omitted non-adiabaticity in the resonant coupling $C_k$. 
In the above notation $q_{ij}<0$ and $C_k>0$ for all $(i,j,k)$. 
From the increase of the nonlinear periods in the case of 
non-resonant hydrodynamical models it follows that the 
$r_k$ parameters are also negative. The frequency mismatch 
is computed from the linear eigenfrequencies as 
$\Delta\omega=\omega_3-2\omega_1$.  

The most general stationary solution of these equations is the 
so-called ``bump'' solution with $A_1\neq0$, $A_2=0$ and $A_3\neq0$. 
In this case we have a single-period pulsation due to phase 
lock and the higher frequency overtone causes a bump appearing,  
e.g., on the radial velocity curve. This solution is thought 
to be the underlying physical model of the bump Cepheids 
(Simon and Schmidt, Buchler and coworkers, see \cite{sim1976}, 
\cite{buc1986}, \cite{bum1990}). 

Yet another stationary solution of the above equations was proposed 
by Dziembowski and Kov\'acs \cite{dzi1984} to model steady 
double-mode pulsation. In this model the resonance between mode `1' 
and `3' leads to the decrease of the amplitude of mode `1' relative 
to its single-mode non-resonant value. This leads to a lower stability 
of the above `bump' solution against perturbations toward mode `2'. 
If the resonance is strong enough ($\Delta\omega$ is small) and some 
other, not too restrictive conditions are also satisfied, mode 
`2' will grow and this will eventually lead to a steady triple-mode 
pulsation (with two `essential' periodicity in the system, since 
mode `1' and `3' remain in resonant frequency synchronization). 
Although the mechanism was shown working also in direct hydrodynamical 
simulations \cite{kov1988}, the required resonance condition 
is not satisfied in standard RR~Lyrae models.\footnote{In the case 
of Cepheids we do have such a resonance but it occurres in the 
bump regime -- near 10~days period -- that is not where most of 
the beat (double-mode) Cepheids can be found. Nevertheless, the 
destabilizing effect of the 2:1 resonance in respect of the 
single-period bump solution was also demonstrated for this class 
of models by hydrodinamical simulations by \cite{bum1990}, \cite{mos1990}.} 

%
%
\medskip
\begin{table}[h]
\begin{tabular}{ll}
\hline
{\bf Code:}   
& Castor/Stellingwerf radiative, LANL opacities \\
{\bf Zoning:} 
& $N_{\rm shell}=100$,  $N_{\rm HIZ} = 30$\tablenote{
Number of mass shells from the surface down to the Hydrogen ionization 
zone at $T=11000$~K.} \\
{\bf Stellar parameters:} 
& $M/M_{\odot}=0.90$, $L/L_{\odot}=40.0$, $T_{\rm eff}=6400$~K \\
{\bf Composition:}
& $X=0.760$, $Z=0.001$ throughout, except in $1.0E5 < T < 1.0E6$, where \\ 
& $X=0.661$, $Y=0.239$, $Z=0.100$, with ramped Z/opacity-enhancement of \\
& $100$ and $20$, respectively. \\  
{\bf Numerical viscosity:} 
& $CQ=4.0$,  $\alpha=0.005$ \\
\hline
\end{tabular}
\caption{Stellar and model construction parameters of the Blazhko toy model.}
\label{table3}
\end{table}  
%

%
%
\medskip
\begin{table}[h]
\begin{tabular}{lrrrrr}
\hline
\tablehead{1}{c}{}{Type}
  & \tablehead{1}{c}{c}{$P_0$}
  & \tablehead{1}{c}{c}{$P_1$}
  & \tablehead{1}{c}{c}{$P_2$}
  & \tablehead{1}{c}{c}{$P_3$}
  & \tablehead{1}{c}{c}{$P_2/P_0$} \\
  & \tablehead{1}{c}{c}{$\eta_0$}
  & \tablehead{1}{c}{c}{$\eta_1$}
  & \tablehead{1}{c}{c}{$\eta_2$}
  & \tablehead{1}{c}{c}{$\eta_3$}
  & \tablehead{1}{c}{c}{$P_1/P_0$} \\
\hline
NOT ENH\tablenote{Standard linear models (no opacity/metal enhancements).}
        & $0.42$ & $0.32$ & $0.26$ & $ 0.21$ & $0.62$ \\
        & $0.01$ & $0.03$ & $0.03$ & $ 0.00$ & $0.76$ \\
ENH\tablenote{Opacity/metal enhanced models as described in the text.}     
        & $0.53$ & $0.33$ & $0.26$ & $ 0.21$ & $0.49$ \\
        & $0.02$ & $0.03$ & $0.04$ & $-0.01$ & $0.62$ \\
\hline
\end{tabular}
\caption{LNA properties of the Blazhko toy model.}
\label{table4}
\end{table}  

In the work of Moskalik \cite{mos1986} two additional important 
simplifications were made in the 2:1 resonance system given by 
Eqs. (13)--(16): (a) the interaction with the non-resonant FO mode 
(Eq. 14) was omitted; (b) saturation in the third mode was 
discarded (i.e., all \{$q_{3i}$\} were set equal to zero), but 
it was assumed that the mode was linearly damped ($\kappa_3<0$). 
With the simplified system the condition of stability of 
the resonant solution was addressed by allowing wide ranges of 
parameters, more importantly those of the frequency mismatch 
$\Delta\omega$ and the damping rate of the resonant overtone 
$\kappa_3$. The parameter survey disclosed a wide range of 
possible limiting states. For high damping rates of mode `3' 
the stable two-mode fixed point solution was the only solution. 
The situation became rather complex when the damping rate was 
lowered (see his Fig.~1). In general, for a very a mild damping 
of mode `3', no stable fixed point was found for very wide a 
range of $\Delta\omega$. Since, for standard RR~Lyrae models, 
the resonant mode is the 3rd overtone and it is not too mildly 
damped, Moskalik investigated this regime in more detail. 
Here, the modulated solution was achieved for non-exact resonances 
($\Delta\omega\neq0$) with some additional conditions put on 
the non-resonant period change terms (parameters \{$r_k$\} in 
Eq.~16). Unfortunately, all these seem to be rather restrictive, 
since the large number of various hydrodynamical simulations 
performed during the past 23~years did not show signs of 
modulated variations. 

In contemplating the disappearance of stable fixed point 
solutions in the mildly damped case, we decided to test the 
case when this happens in real hydrodynamical models. We 
searched for a means to bring one of the overtones in 2:1 
resonance with the fundamental mode and make it less damped 
(or if possible, even excited, since the more physical 
hydrodynamical models do have the saturation terms \{$q_{3i}$\} 
that have been neglected in \cite{mos1986} and they will 
always increase the effective damping in mode `3').    

To create RR~Lyrae models with the above properties, we tried 
the `good old-fashioned' opacity/heavy element enhancement 
\cite{sim1982}, \cite{zdr2008}\, \cite{chr2009}. The required 
size of enhancement is rather large, so there is no good 
physical reason to introduce such change in the models.\footnote{See 
however the `Iron Project' \cite{oel2009} and \cite{mon2008} 
for some renewed efforts on the revisitation of the opacities 
of complex heavy elements.} Our rather vague excuse for doing this 
is the assumption that there was an element enhancement in the 
earlier phase of evolution (e.g., during the ascent to the first 
giant branch, due to the capture of a companion of high heavy 
element content -- see \cite{sie1998}). Somehow mixing was 
avoided and most of the heavy elements settled in the deep 
interior (below the outer convection zone). As a result, they 
do not affect the measured iron content for these typically 
low-metal stars.\footnote{
It is interesting to note that in the Blazhko star TY Gruis 
there are signatures of very large enhancements in some heavy 
elements (see \cite{pre2006}), but this is most probably some 
peculiarity of this object rather than a general feature of the 
Blazhko stars.} Admittedly, this whole scenario is highly unlikely, 
but the present toy model is not intended to be too physical. 

For computing the purely radiative static, linear non-adiabatic (LNA) 
and nonlinear models, we used the same code as in our earlier works 
\cite{buc1990}. For a comfortable scaling of the opacity, we used 
Stellingwerf's \cite{ste1975} analytical formula. For an easier 
overview, in Table~3 we summarized the stellar and model construction 
parameters. We note that the large opacity enhancement is relative to 
the Los Alamos opacities, prior to the OPAL/OP revision of these 
\cite{rog1992}, \cite{sea1993}, \cite{igl1996}. Therefore, in respect 
of the, e.g., OPAL opacities the increase is `only' a factor of five 
or so. In addition to the model presented we computed several/many 
others and the properties to be described below can be regarded as 
generic. 

The LNA properties of the model are shown in Table~4. For comparison, 
we also give the corresponding non-enhanced properties. We see that 
the largest changes due to Z/OP enhancement occurred in the FU mode. 
The drastic change in the period has resulted the reach of near 
2:1 resonance condition between $P_0$ and $P_2$ with a moderate 
increase in the growth rate of the second overtone. These are 
very favorable changes for the anticipated condition of amplitude 
modulation within the resonant model of \cite{mos1986} (not so for 
the standard horizontal branch models, since the values of $P_1/P_0$ 
and $P_0$ are out of range for the double-mode stars and for the 
PLC relations \cite{kov2001}). 

%
%
\begin{figure}[h]
  \includegraphics[height=0.43\textheight]{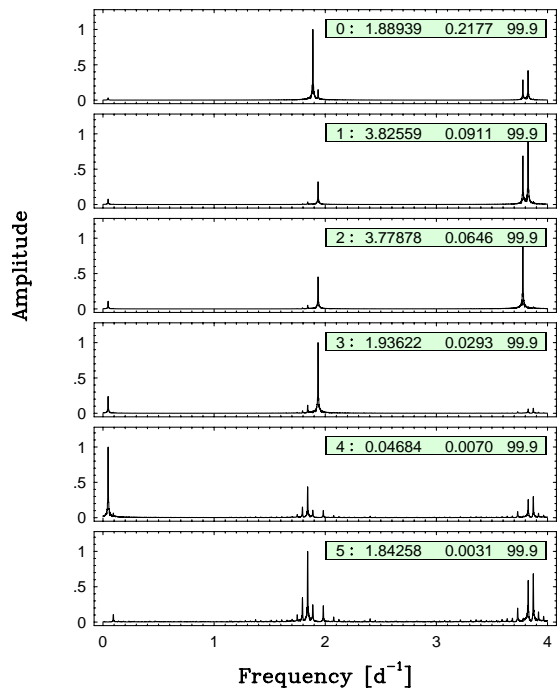}
  \caption{The frequency spectra of the bolometric light 
           curve of the resonant toy model (see Tables~3 
	   and 4). Each spectrum is normalized to the highest 
	   peak. Numbers in the insets of the panels show: 
	   the prewhitening order, the peak frequency 
	   (in [d$^{-1}$]), the corresponding amplitude 
	   (in [mag]) and the signal-to-noise ratio (which 
	   is given as $99.9$, if it is greater than 100).  
	   }
\end{figure}

We initialized the static model with a velocity perturbation 
corresponding to the fundamental mode and integrated in time 
until the asymptotic state was reached. Figure~2 shows the 
frequency spectra of the bolometric light variation of the 
last $500$ $P_0$ cycles. Successive prewhitening of the 
time series revealed the various combination frequencies of 
$\nu_0=1.88939$~d$^{-1}$ and $\nu_m=0.04684$~d$^{-1}$, 
corresponding to the fundamental mode and to the modulation 
frequencies, respectively. It is important to note that the 
components are in very tight numerical relations. Alternatively, 
and perhaps more naturally, we can represent all frequencies 
as linear combinations of $\nu_0$ above and $\nu_2=3.82559$~d$^{-1}$, 
corresponding to the second overtone. This representation is 
very similar to the one advocated by \cite{bor1980} and 
\cite{bre2006}. Indeed, if the asymptotic state is an FU/SO 
double-mode pulsation rather than a genuine amplitude modulation, 
then the higher frequency modulation component at $\nu_0$ is 
$\nu_2-\nu_0$, whereas the much lower amplitude one at the lower 
frequency side corresponds to $3\nu_0-\nu_2$, in line with the 
expected lower amplitude for such a high order combination.  

%
%
\begin{figure}[h]
  \includegraphics[height=.33\textheight]{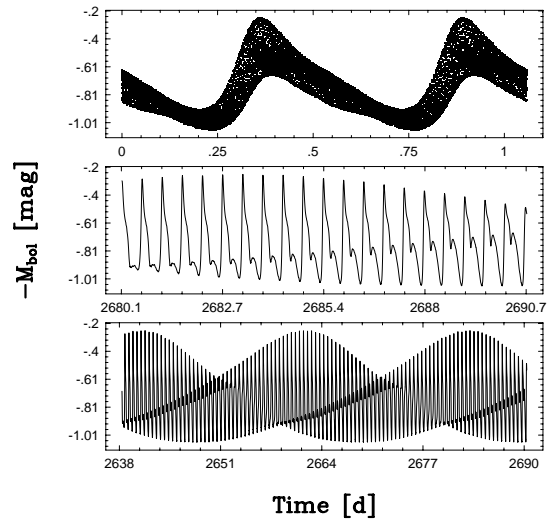}
  \caption{Bolometric light curve of the resonant toy model. 
           Panels from top to bottom show the light curve 
	   folded by the pulsation period, close-up of the 
	   time series and longer stretch of it, covering 
	   three modulation periods.}
\end{figure}

The bolometric light variation is shown in Fig.~3. In the middle 
panel we see the movement of the bump due to the presence of the 
second overtone. The folded light curve exhibits the combination 
of phase and amplitude variations. All these are also observable 
in real Blazhko stars. 

We think that the the above model works the way a toy model should; 
it is certainly entertaining, but requires most probably 
unacceptable torture of the stellar structure. Further temptations 
for additional contemplation of a possible viability of the idea 
is hampered also by the lack of models with nearly symmetric 
modulation components.

%
%
\section{Conclusion and prospects}
In spite of the renewed effort during the past 15 years, so 
far none of the observational and theoretical studies led to 
a breakthrough in our lack of understanding the Blazhko 
phenomenon. More than a thousand RR~Lyrae stars are known in 
various stellar systems that show periodic amplitude- and 
phase-modulations. Importantly, the phenomenon has been 
detected recently also in Cepheids \cite{mos2009} with a 
seemingly quite high incidence rate (at least for some 
classes of Cepheids). There are observational projects 
(Jurcsik and coworkers, Kolenberg and coworkers) that study 
more individual cases and see if there is anything in the 
light and color variations that might give us a clue on the 
underlying mechanism. Sparsely though, but very importantly, 
there are also spectroscopic works aimed at the detection 
of magnetic field in RR~Lyrae stars \cite{cha2004}, 
\cite{kol2009}. Theoretical works \cite{shi2000}, \cite{now2001} 
that targeted the two main (magnetic oblique rotator/pulsator 
and nonradial resonant rotator/pulsator) models have failed 
mostly for being unable to interpret the unequal modulation 
components in the frequency spectra and also for other 
observational and theoretical reasons (strong nonlinearity 
of the nonradial modes and apparent absence of strong magnetic 
field).  
    
It seems to be that the inspiration for new types of models 
should come from the observational side. Although current 
ground-based observations can also target searches for 
additional frequency components in the light variation in 
the millimagnitude level, the real breakthrough will come from 
the current space projects CoRoT and Kepler. First signs 
of this event have already been witnessed from the preliminary 
results coming from CoRoT (Chadid et al., these Proceedings). 
Whether the appearance of additional small amplitude components 
will be a common feature of all Blazhko stars or this feature 
remains to be a rare occurrence, is still to be seen. Obviously, 
the presence of low-amplitude modulated group of frequencies 
would be suggestive of a so far not investigated effect of 
the dynamical interaction of several/many nonradial and radial 
modes \cite{dzi2004}. It is also interesting how the incidence 
rate of the Blazhko phenomenon will change as the signal-to-noise 
ratio of the measurements increases. There are already signs that 
the phenomenon is much more common than we thought (Jurcsik, these 
Proceedings). Unfortunately, the number of the spectroscopic works 
is far below that of the photometric ones. We think that line 
profile analyses (even if they were based on snapshot observations) 
would be very important for future modeling. An upper limit on 
the size of the nonradial component of the velocity field would 
be instrumental in considering nonradial modes either as members 
of a periodic energy sink, or retain their leading role in 
establishing the amplitude modulation, or reject the nonradial 
mode interaction altogether and search for other mechanisms.

%
%
\begin{theacknowledgments}
  We highly appreciate the fruitful correspondence with George Preston. 
  Lively discussions during the meeting with Wojtek Dziembowski, 
  Robert Buchler, Katrien Kolenberg and Merieme Chadid were very 
  stimulating. Correspondence with Radek Smolec on nonlinear period 
  change was very helpful. Additional discussions with J\'ozsef Benk\H o 
  and R\'obert Szab\'o on the CoRoT RR~Lyrae are also thanked. Comments 
  on the manuscript by Istv\'an D\'ek\'any were very instrumental 
  during the final phase of writing up. We acknowledge the support 
  of the Hungarian Research Fund (OTKA) K-60750. 
\end{theacknowledgments}

%
%

\end{document}